\documentstyle[12pt]{article}

\newcommand{\beq}{\begin{equation}}
\newcommand{\eeq}{\end{equation}}
\newcommand{\bea}{\begin{eqnarray}}
\newcommand{\eea}{\end{eqnarray}}
\newcommand{\gf}{G_{\mbox{{\scriptsize F}}}}
\newcommand{\heff}{H_{\mbox{{\scriptsize eff}}}}
\newcommand{\xx}{(X_1\,X_2)_f}
\newcommand{\xxc}{(X_1\,X_2)_f^{\mbox{{\tiny C}}}}
\newcommand{\bxx}{B_s\to X_1\,X_2}
\newcommand{\bxxbar}{\overline{B_s}\to X_1\,X_2}
\newcommand{\bxxc}{B_s\to (X_1\,X_2)^{\mbox{{\tiny C}}}}
\newcommand{\bxxcbar}{\overline{B_s}\to
(X_1\,X_2)^{\mbox{{\tiny C}}}}
\newcommand{\bsbar}{\overline{B_s}}
\newcommand{\lab}{\label}
\newcommand{\order}{{\cal O}}
\newcommand{\aftilde}{A_{\tilde f}^\ast}
\newcommand{\aftildebar}{\overline{A}_{\tilde f}^\ast}
\newcommand{\afc}{A_f^{\mbox{{\tiny C}}}}
\newcommand{\afcbar}{\overline{A}_f^{\mbox{{\tiny C}}}}
\newcommand{\afctilde}{A_{\tilde f}^{\mbox{{\tiny C}}\ast}}
\newcommand{\afctildebar}{\overline{A}_{\tilde f}^{\mbox{{\tiny
C}}\ast}}
\newcommand{\af}{A_f}
\newcommand{\afbar}{\overline{A}_f}
\newcommand{\etap}{\eta_{\mbox{{\tiny P}}}^f}
\newcommand{\etaptilde}{\eta_{\mbox{{\tiny P}}}^{\tilde f}}
\newcommand{\lambdaf}{\lambda_f}
\newcommand{\lambdaftilde}{\lambda_{\tilde f}}
\newcommand{\lambdacf}{\lambda_f^{\mbox{{\tiny C}}}}
\newcommand{\lambdacftilde}{\lambda_{\tilde f}^{\mbox{{\tiny C}}\ast}}
\newcommand{\phicp}{\phi_{\mbox{{\tiny CP}}}(B_s)}
\newcommand{\cp}{({\cal CP})}
\newcommand{\oabar}{\overline{O}_1}
\newcommand{\obbar}{\overline{O}_2}
\newcommand{\mfbar}{\overline{M}_f}
\newcommand{\mf}{M_f}
\newcommand{\mftilde}{M_{\tilde f}^\ast}
\newcommand{\bstack}{\stackrel{{\mbox{\tiny (---)}}}{B_s}}
\newcommand{\dstacko}{\stackrel{{\mbox{\tiny (---)}}}{D^0}}
\newcommand{\dstackoast}{\stackrel{{\mbox{\tiny (---)}}}{D^{\ast0}}}
\newcommand{\dstackres}{\stackrel{{\mbox{\tiny (---)}}}{D_1}(2420)^0}
\newcommand{\dstackresres}{\stackrel{{\mbox{\tiny (---)}}}{D^{\ast\ast0}}}

\begin{document}

\setcounter{page}{0}
\thispagestyle{empty}

\rightline{TTP96-08}
\rightline{FERMILAB-PUB-96/080-T}
\rightline{hep-ph/9605221}
\rightline{May 1996}
\bigskip
\begin{center}
{\Large {\bf CP violation and the CKM angle $\gamma$ from}}
\end{center}
\begin{center}
{\Large{\bf angular distributions of untagged $B_s$ decays}}
\end{center}
\begin{center}
{\Large{\bf governed by $\bar b\to\bar c u\bar s$}}\\
\end{center}
\vspace{0.2cm}
\smallskip
\begin{center}
{\large{\sc Robert Fleischer}}\footnote{Internet: {\tt
rf@ttpux1.physik.uni-karlsruhe.de}}\\
{\sl Institut f\"{u}r Theoretische Teilchenphysik}\\
{\sl Universit\"{a}t Karlsruhe}\\
{\sl D--76128 Karlsruhe, Germany}\\
\vspace{0.7cm}
{\large{\sc Isard Dunietz}}\footnote{Internet: {\tt
dunietz@fnth15.fnal.gov}}\\
{\sl Theoretical Physics Division}\\
{\sl Fermi National Accelerator Laboratory}\\
{\sl Batavia, IL 60510, USA}\\
\vspace{0.8cm}
{\large{\bf Abstract}}\\
\end{center}
\vspace{0.1cm}
We demonstrate that time-dependent studies of angular distributions
for $B_s$ decays caused by $\bar b\to\bar cu\bar s$ quark-level 
transitions extract {\it cleanly} and {\it model-independently} the CKM 
angle $\gamma$. This CKM angle could be cleanly determined from 
{\it untagged} $B_s$ decays alone, if the lifetime difference between 
the $B_s$ mass eigenstates $B_s^L$ and $B_s^H$ is sizable. The 
time-dependences for the relevant {\it tagged} and {\it untagged} 
observables are given both in a general notation and in terms of linear 
polarization states and should exhibit large CP-violating effects. 
These observables may furthermore provide insights into the hadronization 
dynamics of the corresponding exclusive $B_s$ decays thereby allowing 
tests of the factorization hypothesis.

\newpage

\section{Introduction}\lab{intro}

Some time ago it was pointed out in \cite{adk} that a clean
measurement of the angle $\gamma$ of the usual ``non-squashed''
unitarity triangle \cite{ut} of the Cabibbo--Kobayashi--Maskawa
matrix (CKM matrix) \cite{km} is possible by studying the time
dependence of the color-allowed decays $\bstack\to D_s^\pm K^\mp$.
A similar analysis of the color-suppressed modes $\bstack\to\dstacko\phi$
provides in principle also clean information about $\gamma$ \cite{aklgam}.
Because current detectors have difficulties in observing the soft photon 
in $D_s^* \to D_s \gamma$ decays, Aleksan, Le Yaouanc, Oliver, Pene and 
Raynal employed several plausible assumptions to show that the CKM angle 
$\gamma$ can still be extracted from partially reconstructed $B_s$ modes 
where that soft photon could be missed~\cite{aleksanfrench}.

Unfortunately in all these strategies {\it tagging}, i.e.\ the
distinction between initially present $B_s$ and $\bsbar$ mesons,
is essential. Moreover one has to resolve the rapid
$B_s-\bsbar$ oscillations, which may arise from the expected large
mass difference $\Delta m\equiv m_H-m_L>0$ between the mass
eigenstates $B_s^H$ (``heavy'') and $B_s^L$ (``light'') \cite{smexpt}.
This is a formidable experimental task. In a recent paper
\cite{dunietz} these methods have been re-considered
in light of the expected perceptible lifetime difference
\cite{deltagamma} between $B_s^H$ and $B_s^L$. There it has
been shown that the rapid $\Delta mt$--oscillations cancel in
{\it untagged} data samples.
Whereas the extraction of $\gamma$ from untagged
$\stackrel{(-)}{B_s} \rightarrow D^\pm_s K^\mp$ requires some mild
additional theoretical input, it does not require any theory beyond the
validity of the CKM model from untagged
$\stackrel{(-)}{B_s} \rightarrow \stackrel{(-)}{D^0}
\phi$  decays \cite{dunietz}.

In a recent publication \cite{df1} we have investigated quasi two
body modes $\bxx$ into admixtures of different CP eigenstates
where both $X_1$ and $X_2$ carry spin and continue to decay through
CP-conserving interactions. The time-dependent angular distributions
for the {\it untagged} decays $B_s\to D_s^{\ast +}\,D_s^{\ast-}$
and $B_s\to J/\psi\,\phi$ determine the Wolfenstein parameter $\eta$
\cite{wolf}. If one uses $|V_{ub}|/|V_{cb}|$ as an additional
input, the CKM angle $\gamma$ can be fixed. That input allows, however,
also the determination of $\eta$ (or $\gamma$) from the mixing-induced
CP asymmetry of $B_d\to J/\psi\,K_S$ measuring $\sin 2\beta$ ($\beta$ is
another angle of the unitarity triangle \cite{ut}). Comparing these
two results for $\eta$ (or $\gamma$) obtained from $B_s$ and $B_d$
modes, respectively, an interesting test whether the $B_s-\overline{B_s}$
and $B_d-\overline{B_d}$ mixing phases are described by the Standard Model
or receive additional contributions from ``New Physics'' can be
preformed. Another application of the formalism developed in \cite{df1}
is the point that a determination of $\gamma$ is possible by using the
$SU(2)$ isospin symmetry of strong interactions to relate {\it untagged}
data samples of $B_s\to K^{\ast+}\,K^{\ast-}$ and $B_s\to K^{\ast0}\,
\overline{K^{\ast0}}$.

\vspace{0.2cm}

Having all these results in mind it is quite natural to ask what can
be learned from time-dependent {\it untagged} measurements of the
angular distributions for $\bstack\to D_s^{\ast\pm}\,K^{\ast\mp}$,
$D_{s1}(2536)^\pm K^{\ast\mp}$, $D_s^{\ast\ast\pm}K^{\ast\mp}$
and $\bstack\to\dstackoast\phi$, $\dstackres\phi$,
$\dstackresres\phi$ or -- more generally -- from $B_s$
modes governed by $\bar b\to\bar cu\bar s$ quark-level transitions.
Since the photon(s) in the strong or electromagnetic decays of $D^*_s$
and $D^{\ast0}$ are more difficult to detect than charged particles for
generic detectors, we listed also higher resonances because of
their significant all-charged final states, such as $D_{s1} (2536)^+
\rightarrow D^{*+} K^0 ,D^{*+} \rightarrow \pi^+ D^0$ or $D_1 (2420)^0
\rightarrow D^{*+} \pi^- ,D^{*+} \rightarrow \pi^+ D^0$.
The $K^{*\mp}$ in the above $B_s$-decays can be substituted by either a
strange resonance or a collection of strange resonances with common spin
and parity quantum numbers.

While our note focusses on quasi two body modes where each body has a
well-defined spin and parity, a complementary report discusses the
effects when several resonances contribute to the final state
\cite{ads}. In the former case the final states cannot be classified
by their CP eigenvalues as in \cite{df1}. However, they can instead
be classified by their parities. To this end linear polarization
states \cite{rosner} are particularly useful. As we will demonstrate in
the present paper, the {\it untagged} angular distributions for
such $B_s$ decays may inform us in a clean way
about $\gamma$, if the lifetime difference between $B_s^H$ and $B_s^L$
is in fact sizable. In particular we do not need any
theoretical input to extract this quantity from the
{\it untagged} data samples which exhibit in addition interesting
CP-violating effects. Furthermore essentially the whole
hadronization dynamics can be extracted from these angular correlations.
Since, for example, the $\bstack\to D_s^{\ast\pm}\,K^{\ast\mp}$,
$D_{s1}(2536)^\pm K^{\ast\mp}$, $D_s^{\ast\ast\pm}K^{\ast\mp}$
modes are color-allowed whereas the $\bstack\to\dstackoast\phi$,
$\dstackres\phi$, $\dstackresres\phi$
channels are color-suppressed, the factorization hypothesis
\cite{fact1,fact2}, which has some justification within the
$1/N_{\mbox{{\scriptsize C}}}$--expansion \cite{bgr}, should work
quite well in the former case and should be very
questionable in the latter case \cite{bjorken}. Therefore we expect
significant non-factorizable contributions to the angular distributions
for the $\bstack\to\dstackoast\phi$, $\dstackres\phi$, $\dstackresres\phi$
decays. The explicit angular distributions for some of these decays will be
given in a separate publication \cite{angdistr}. There also appropriate
weighting functions are given allowing an efficient extraction of the
corresponding observables from experimental data with the help of a
{\it moment analysis} (see \cite{dqstl,ddf1}).

Our paper is organized as follows: The time-dependences of the
observables of the angular distributions are calculated in
Section~\ref{evol} in terms of a general notation that allows an
easy comparison with the results presented in \cite{df1}. In
Section~\ref{linpol} these time-dependences are given in terms of
linear polarization states which provide a useful tool to calculate the
explicit angular distributions for final state configurations having
definite parities. There we demonstrate explicitly that the observables
of the {\it untagged} angular distributions for the $\bar b\to\bar cu\bar
s$ (and $\bar b\to\bar uc\bar s$) decays suffice to extract the CKM angle
$\gamma$. The issue of CP violation in untagged data samples is
discussed in Section~\ref{cpviol} and the main results of our paper are
summarized briefly in Section~\ref{sum}.

\section{Calculation of the general time-evolutions}\lab{evol}

In the case of the decays considered in this paper, the transition
amplitudes for the quasi two body modes $\bxxbar$ and $\bxx$ can
be expressed as hadronic matrix elements of low energy effective
Hamiltonians having the following structures:
\bea
\heff\,(\bxxbar)&=&\frac{\gf}{\sqrt{2}}\overline{v}\left[C_1(\mu)\oabar+
C_2(\mu)\obbar\right]\lab{e1}\\
\heff\,(\bxx)&=&\frac{\gf}{\sqrt{2}}v^\ast\left[C_1(\mu)O_1^\dagger+
C_2(\mu)O_2^\dagger\right],\lab{e2}
\eea
where $\overline{v}$ and $v$ denote appropriate CKM factors,
$\overline{O}_k$ and $O_k$ $(k\in\{1,2\})$ are four-quark
operators (``current-current'' operators in our case) and
$C_1(\mu)$ and $C_2(\mu)$ are the Wilson coefficient functions
of these operators. They can be calculated perturbatively and
contain the whole short distance dynamics. As usual
$\mu=\order(m_b)$ is a renormalization scale. To be definite, for
$X_1\,X_2\in\{D_s^{\ast+}\,K^{\ast-}$, $D_{s1}(2536)^+K^{\ast-}$,
$D_{s}^{\ast\ast+}K^{\ast-}$, $D^{\ast0}\,\phi$, $D_1(2420)^0\phi$,
$D^{\ast\ast0}\phi\}$ we have
\beq\lab{e3}
\begin{array}{rcl}
\oabar&=&\left(\bar s_\alpha u_\beta\right)_{\mbox{{\scriptsize V--A}}}
\left(\bar c_\beta b_\alpha\right)_{\mbox{{\scriptsize V--A}}}\\
\obbar&=&\left(\bar s_\alpha u_\alpha\right)_{\mbox{{\scriptsize V--A}}}
\left(\bar c_\beta b_\beta\right)_{\mbox{{\scriptsize V--A}}}
\end{array}
\eeq
\beq\lab{e4}
\begin{array}{rcl}
O_1&=&\left(\bar s_\alpha c_\beta\right)_{\mbox{{\scriptsize V--A}}}
\left(\bar u_\beta b_\alpha\right)_{\mbox{{\scriptsize V--A}}}\\
O_2&=&\left(\bar s_\alpha c_\alpha\right)_{\mbox{{\scriptsize V--A}}}
\left(\bar u_\beta b_\beta\right)_{\mbox{{\scriptsize V--A}}},
\end{array}
\eeq
where the greek indices denote $SU(3)_{\mbox{{\scriptsize C}}}$
color indices, and the CKM factors are given by
\beq\lab{e5}
\begin{array}{rcl}
\overline{v}&=&V_{us}^\ast V_{cb}\\
v&=&V_{cs}^\ast V_{ub}.
\end{array}
\eeq
Nowadays the Wilson coefficients $C_1(\mu)$ and $C_2(\mu)$ are available
beyond the leading logarithmic approximation \cite{acmp,bw}. A
nice review of such next-to-leading order calculations has been given
recently in \cite{burasnlo}, and we refer the reader to that publication
for the details of such calculations.

Applying a similar notation as in \cite{df1}, we obtain the following
transition amplitudes for decays of $B_s$ and $\bsbar$ mesons into a
configuration $f$ of the quasi two body state $X_1\,X_2$, where $f$
is a label that defines the relative polarizations of the two hadrons
$X_1$ and $X_2$:
\bea
\afbar&\equiv&\langle\xx|\heff\,(\bxxbar)|\bsbar\rangle=\frac{\gf}{\sqrt{2}}
\,\overline{v}\,\mfbar\lab{e6}\\
\af&\equiv&\langle\xx|\heff\,(\bxx)|B_s\rangle=\etap\, e^{i\phicp}\,
\frac{\gf}{\sqrt{2}}\, v^\ast\, M_f\lab{e7}
\eea
with
\bea
\mfbar&\equiv&\langle\xx|C_1(\mu)\oabar+C_2(\mu)\obbar|\bsbar\rangle
\lab{e8}\\
M_f&\equiv&\langle\xxc|C_1(\mu)O_1+C_2(\mu)O_2|\bsbar\rangle.\lab{e9}
\eea
In order to evaluate (\ref{e7}) we have performed the CP transformations
\bea
\lefteqn{\langle\xx|C_1(\mu)O_1^\dagger+C_2(\mu)O_2^\dagger|B_s\rangle}
\nonumber\\
&&=\langle\xx|\cp^\dagger\cp\left[C_1(\mu)O_1^\dagger+C_2(\mu)O_2^\dagger
\right]\cp^\dagger\cp|B_s\rangle\lab{e10}\\
&&=\etap\,e^{i\phicp}\,\langle\xxc|C_1(\mu)O_1+C_2(\mu)O_2|\bsbar\rangle
\nonumber
\eea
by taking into account the relations
\beq\lab{e11}
\cp O_k^\dagger\cp^\dagger=O_k
\eeq
and
\beq\lab{e12}
\cp|B_s\rangle=e^{i\phicp}\,|\bsbar\rangle
\eeq
\beq\lab{e13}
\cp|\xx\rangle=\etap\,|\xxc\rangle.
\eeq
Here $\phicp$ parametrizes the applied CP phase convention and
$\etap\in\{-1,+1\}$ denotes the parity eigenvalues of the configurations
$f$ of $X_1\,X_2$. In terms of linear polarization amplitudes
\cite{rosner} (see also \cite{ddlr}) we have $\eta_{\mbox{{\tiny P}}}^0=
\eta_{\mbox{{\tiny P}}}^\parallel=+1$ and $\eta_{\mbox{{\tiny
P}}}^\perp=-1$ for $X_1\,X_2\in\{D_s^{\ast+}K^{\ast-}, D^{\ast0}\phi\}$.
In contrast, for $X_1\,X_2\in\{D_{s1}(2536)^+ K^{\ast-}, D_1(2420)^0\phi\}$ 
we have $\eta_{\mbox{{\tiny P}}}^0=\eta_{\mbox{{\tiny P}}}^\parallel=-1$ 
and $\eta_{\mbox{{\tiny P}}}^\perp=+1$.

Let us now consider the $\bsbar$ and $B_s$ decays into the charge-conjugate
quasi two body states $(X_1\,X_2)^{\mbox{{\tiny C}}}$. In the case
relevant for the present paper corresponding to
$X_1\,X_2\in\{D_s^{\ast+}\,K^{\ast-}$, $D_{s1}(2536)^+K^{\ast-}$,
$D_{s}^{\ast\ast+}K^{\ast-}$, $D^{\ast0}\,\phi, D_1(2420)^0\phi$,
$D^{\ast\ast0}\phi\}$ we have
$(X_1\,X_2)^{\mbox{{\tiny C}}}\in\{D_s^{\ast-}\,K^{\ast+}$,
$D_{s1}(2536)^-K^{\ast+}$, $D_{s}^{\ast\ast-}K^{\ast+}$,
$\overline{D}^{\ast0}\,\phi$, $\overline{D}_1(2420)^0\phi$,
$\overline{D}^{\ast\ast0}\phi\}$, respectively. If the charge-conjugate
states are present in a configuration $f$ with parity eigenvalue $\etap$,
a similar calculation as sketched above yields
\bea
\afcbar&\equiv&\langle\xxc|\heff\,(\bxxcbar)|\bsbar\rangle=
\frac{\gf}{\sqrt{2}}\,v\,M_f\lab{e14}\\
\afc&\equiv&\langle\xxc|\heff\,(\bxxc)|B_s\rangle=\etap\,
e^{i\phicp}\,\frac{\gf}{\sqrt{2}}\,\overline{v}^\ast\,\mfbar.\lab{e15}
\eea
Using these results and the well-known formalism describing $B_s-\bsbar$
mixing \cite{dunietz,dr}, we obtain the following expressions for
initially, i.e.\ at $t=0$, present $B_s$ and $\bsbar$ mesons:
\bea
\lefteqn{\aftilde(t)\,\af(t)=\frac{\gf^2}{2}\,|v|^2\,\etaptilde\,\etap\,
\mftilde\,\mf}
\label{e16}\\
&&\times\left[|g_+(t)|^2+\etaptilde\,\lambdaftilde^\ast\,
g_+(t)\,g_-^\ast(t)+\etap\,\lambdaf\, g_+^\ast(t)\,g_-(t)+
\etaptilde\,\etap\,\lambdaftilde^\ast\,
\lambdaf\,|g_-(t)|^2\right]\nonumber
\eea
\bea
\lefteqn{\aftildebar(t)\,\afbar(t)=\frac{\gf^2}{2}\,|v|^2\,
\etaptilde\,\etap\,\mftilde\,\mf}
\label{e17}\\
&&\times\left[|g_-(t)|^2+\etaptilde\,\lambdaftilde^\ast\,
g_+^\ast(t)\,g_-(t)+\etap\,\lambdaf\, g_+(t)\,g_-^\ast(t)+
\etaptilde\,\etap\,\lambdaftilde^\ast\,
\lambdaf\,|g_+(t)|^2\right]\nonumber
\eea
\bea
\lefteqn{\afctilde(t)\,\afc(t)=\frac{\gf^2}{2}\,|v|^2\,\mftilde\,\mf}
\label{e18}\\
&&\times\left[|g_-(t)|^2+\etaptilde\,\lambdacftilde\,
g_+^\ast(t)\,g_-(t)+\etap\,\lambdacf\, g_+(t)\,g_-^\ast(t)+
\etaptilde\,\etap\,\lambdacftilde\,
\lambdacf\,|g_+(t)|^2\right]\nonumber
\eea
\bea
\lefteqn{\afctildebar(t)\,\afcbar(t)=\frac{\gf^2}{2}\,|v|^2\,\mftilde\,\mf}
\label{e19}\\
&&\times\left[|g_+(t)|^2+\etaptilde\,\lambdacftilde\,
g_+(t)\,g_-^\ast(t)+\etap\,\lambdacf\, g_+^\ast(t)\,g_-(t)+
\etaptilde\,\etap\,\lambdacftilde\,
\lambdacf\,|g_-(t)|^2\right],\nonumber
\eea
where
\bea
|g_\pm(t)|^2&=&\frac{1}{4}\left[e^{-\Gamma_Lt}+e^{-\Gamma_Ht}\pm
2e^{-\Gamma t}\cos(\Delta mt)\right]\lab{e6a}\\
g_+(t)g_-^\ast(t)&=&\frac{1}{4}\left[e^{-\Gamma_Lt}-e^{-\Gamma_Ht}-
2ie^{-\Gamma t}\sin(\Delta mt)\right]\lab{e6b}
\eea
with $\Gamma\equiv(\Gamma_L+\Gamma_H)/2$. The observable $\lambdaf$
is defined through
\beq\lab{e20}
\lambdaf\equiv-\,\etap\,e^{-i\Theta_{M_{12}}^{(s)}}\frac{\afbar}{\af}
\eeq
with
\beq\lab{e21}
\Theta_{M_{12}}^{(s)}=\pi+2\,\mbox{arg}(V_{ts}^\ast V_{tb})-\phicp
\eeq
denoting the phase of the off-diagonal element of the $B_s-\bsbar$
mass matrix. Combining (\ref{e20}) with (\ref{e21}) and (\ref{e6})
and (\ref{e7}), we observe explicitly that the convention dependent
phases $\phicp$ cancel (as they have to!) and arrive at
\beq\lab{e22}
\lambdaf=\exp\left(-2\,i\,\mbox{arg}\{V_{ts}^\ast V_{tb}\}\right)\,
\frac{\overline{v}}{v^\ast}\,\frac{\mfbar}{M_f}.
\eeq
Correspondingly we have introduced
\beq\lab{e23}
\lambdacf\equiv-\frac{1}{\etap\,e^{-i\Theta_{M_{12}}^{(s)}}}\,
\frac{\afc}{\afcbar}=
\left[\exp\left(-2\,i\,\mbox{arg}\{V_{ts}^\ast V_{tb}\}\right)\,
\frac{\overline{v}}{v^\ast}\right]^\ast\frac{\mfbar}{M_f}.
\eeq
Note that $\lambdaftilde$ and $\lambdaftilde^{\mbox{{\tiny C}}}$
can be obtained easily from (\ref{e22}) and (\ref{e23}) by replacing
$f$ with $\tilde f$.

Real or imaginary parts of
bilinear combinations of decay amplitudes like those given in
(\ref{e16})-(\ref{e19}) govern the angular distributions for the
decay products of $X_1$ and $X_2$. In this paper we are focussing on
{\it untagged} angular distributions, where one does not distinguish
between initially present $B_s$ and $\bsbar$ mesons. The corresponding
observables for $\bstack\to X_1\,X_2$ and $\bstack\to
(X_1\,X_2)^{\mbox{{\tiny C}}}$ are related to real or imaginary
parts of
\bea
\lefteqn{\left[\aftilde(t)\,\af(t)\right]\equiv\aftildebar(t)\,\afbar(t)+
\aftilde(t)\,\af(t)=\frac{\gf^2}{4}\,|v|^2\,\etaptilde\,\etap\,\mftilde\,
\mf}\lab{e24}\\
&&\times\left[\left(1+\etaptilde\,\etap\lambdaftilde^\ast\,\lambdaf
\right)\left(e^{-\Gamma_Lt}+e^{-\Gamma_Ht}\right)+\left(\etaptilde\,
\lambdaftilde^\ast+\etap\,\lambdaf\right)\left(e^{-\Gamma_Lt}-
e^{-\Gamma_Ht}\right)\right]\nonumber
\eea
and
\bea
\lefteqn{\left[\afctilde(t)\,\afc(t)\right]\equiv\afctildebar(t)\,\afcbar(t)+
\afctilde(t)\,\afc(t)=\frac{\gf^2}{4}\,|v|^2\,\mftilde\,\mf}\lab{e25}\\
&&\times\left[\left(1+\etaptilde\,\etap\lambdacftilde\,\lambdacf
\right)\left(e^{-\Gamma_Lt}+e^{-\Gamma_Ht}\right)+\left(\etaptilde\,
\lambdacftilde+\etap\,\lambdacf\right)\left(e^{-\Gamma_Lt}-
e^{-\Gamma_Ht}\right)\right],\nonumber
\eea
respectively. In order to evaluate these equations we have combined
(\ref{e16})-(\ref{e19}) with the explicit time-dependences of (\ref{e6a})
and (\ref{e6b}). At present such {\it untagged} studies are obviously much
more efficient from an experimental point of view than tagged analyses.
In the distant future it will be feasible to collect also
{\it tagged} data samples of $B_s$ decays and to resolve the rapid
oscillatory $\Delta mt$--terms. The corresponding tagged observables are
given in (\ref{e16})-(\ref{e19}).

Let us after these general considerations become more specific
in the following section. There we give the time-evolutions
in terms of linear polarization states and demonstrate that the
{\it untagged} observables evolving as real or imaginary parts of
(\ref{e24}) and (\ref{e25}) suffice to extract the CKM angle $\gamma$.

\section{The extraction of the CKM angle $\gamma$}\lab{linpol}

Since it is convenient to give the angular distributions in
terms of the linear polarization states $f\in\{\,0\,,\,
\parallel\,,\,\perp\,\}$ (see \mbox{\cite{rosner,ddlr}}),
let us summarize the corresponding
time-dependences in this section. The linear polarization states are
characterized by the parity eigenvalues $\etap$. If we introduce the
quantity
\beq\lab{erf}
R_f\equiv|R_f|e^{i\rho_f}\equiv\frac{|\overline{v}|}{|v|}
\frac{\mfbar}{M_f},
\eeq
where $\rho_f$ is a CP-conserving strong phase originating from strong
final state interaction processes, we have in our specific case
$X_1\,X_2\in\{D_s^{\ast+}K^{\ast-}$, $D_{s1}(2536)^+K^{\ast-}$,
$D_s^{\ast\ast+}K^{\ast-}$, $D^{\ast0}\phi$, $D_1(2420)^0\phi$,
$D^{\ast\ast0}\phi\}$
\bea
\lambda_f&=&e^{-i\gamma}R_f\\
\lambda_f^{\mbox{{\tiny C}}}&=&e^{+i\gamma}R_f,
\eea
where $\gamma$ is the notoriously difficult to measure CKM angle of the
unitarity triangle \cite{ut}. Using (\ref{e5}) and the Wolfenstein
expansion \cite{wolf} of the CKM matrix by neglecting terms of
$\order(\lambda^2)$, where $\lambda=\sin\theta_{\mbox{{\scriptsize
C}}}=0.22$ is related to the Cabibbo angle, we obtain
\beq
R_f=\frac{1}{R_b}\frac{\mfbar}{M_f}
\eeq
with
\beq
R_b\equiv\frac{1}{\lambda}\frac{|V_{ub}|}{|V_{cb}|}.
\eeq
The CKM factor $R_b$ is constrained by present experimental data to
lie within the range $R_b=0.36\pm0.08$ \cite{cleo,blo,al}.

If we express the hadronic matrix elements $M_f$ defined by (\ref{e9})
in the form
\beq
M_f=|M_f|e^{i\vartheta_f},
\eeq
where $\vartheta_f$ denotes a CP-conserving strong phase shift, the
time-dependent {\it untagged} observables corresponding to the linear
polarization states \cite{rosner} are in the case of $\bstack\to X_1\,X_2$
given by
\bea
\lefteqn{\left[|A_0(t)|^2\right]=\frac{\gf^2}{4}|V_{ub}
V_{cs}|^2|M_0|^2}\lab{e34}\\
&&\times\left[\left(1+|R_0|^2
\right)\left(e^{-\Gamma_Lt}+e^{-\Gamma_Ht}\right)+2|R_0|\cos(\rho_0-\gamma)
\left(e^{-\Gamma_Lt}-e^{-\Gamma_Ht}\right)\right]\nonumber
\eea
\bea
\lefteqn{\left[|A_\parallel(t)|^2\right]=\frac{\gf^2}{4}|V_{ub}V_{cs}|^2
|M_\parallel|^2}\lab{e35}\\
&&\times\left[\left(1+|R_\parallel|^2
\right)\left(e^{-\Gamma_Lt}+e^{-\Gamma_Ht}\right)+2|R_\parallel|
\cos(\rho_\parallel-\gamma)\left(e^{-\Gamma_Lt}-e^{-\Gamma_Ht}\right)
\right]\nonumber
\eea
\bea
\lefteqn{\left[|A_\perp(t)|^2\right]=
\frac{\gf^2}{4}|V_{ub}V_{cs}|^2|M_\perp|^2}\lab{e36}\\
&&\times\left[\left(1+|R_\perp|^2\right)\left(e^{-\Gamma_Lt}+
e^{-\Gamma_Ht}\right)-2|R_\perp|\cos(\rho_\perp-\gamma)
\left(e^{-\Gamma_Lt}-e^{-\Gamma_Ht}\right)\right]\nonumber
\eea
\bea
\lefteqn{\left[A_0^\ast(t)A_\parallel(t)\right]=
\frac{\gf^2}{4}|V_{ub}V_{cs}|^2|M_0||M_\parallel|e^{i(\vartheta_\parallel-
\vartheta_0)}\left[\left(1+|R_0||R_\parallel|e^{i(\rho_\parallel-\rho_0)}
\right)\right.}\lab{e37}\\
&&\left.\times\left(e^{-\Gamma_Lt}+e^{-\Gamma_Ht}\right)+\left(
|R_0|e^{i(\gamma-\rho_0)}+|R_\parallel|e^{-i(\gamma-\rho_\parallel)}\right)
\left(e^{-\Gamma_Lt}-e^{-\Gamma_Ht}\right)\right]\nonumber
\eea
\bea
\lefteqn{\left[A_\parallel^\ast(t)A_\perp(t)\right]=
-\frac{\gf^2}{4}|V_{ub}V_{cs}|^2|M_\parallel||M_\perp|
e^{i(\vartheta_\perp-\vartheta_\parallel)}\left[\left
(1-|R_\parallel||R_\perp|e^{i(\rho_\perp-\rho_\parallel)}
\right)\right.}\\
&&\left.\times\left(e^{-\Gamma_Lt}+e^{-\Gamma_Ht}\right)+\left(
|R_\parallel|e^{i(\gamma-\rho_\parallel)}-|R_\perp|e^{-i(\gamma-
\rho_\perp)}\right)\left(e^{-\Gamma_Lt}-e^{-\Gamma_Ht}\right)\right]\nonumber
\eea
\bea
\lefteqn{\left[A_0^\ast(t)A_\perp(t)\right]=
-\frac{\gf^2}{4}|V_{ub}V_{cs}|^2|M_0||M_\perp|
e^{i(\vartheta_\perp-\vartheta_0)}\left[\left
(1-|R_0||R_\perp|e^{i(\rho_\perp-\rho_0)}
\right)\right.}\\
&&\left.\times\left(e^{-\Gamma_Lt}+e^{-\Gamma_Ht}\right)+\left(
|R_0|e^{i(\gamma-\rho_0)}-|R_\perp|e^{-i(\gamma-
\rho_\perp)}\right)\left(e^{-\Gamma_Lt}-
e^{-\Gamma_Ht}\right)\right].\nonumber
\eea
For the untagged decays into the charge conjugate two body states
we obtain on the other hand the following expressions:
\bea
\lefteqn{\left[|A_0^{\mbox{{\tiny C}}}(t)|^2\right]=
\frac{\gf^2}{4}|V_{ub}V_{cs}|^2|M_0|^2}\lab{e40}\\
&&\times\left[\left(1+|R_0|^2
\right)\left(e^{-\Gamma_Lt}+e^{-\Gamma_Ht}\right)+2|R_0|\cos(\rho_0+\gamma)
\left(e^{-\Gamma_Lt}-e^{-\Gamma_Ht}\right)\right]\nonumber
\eea
\bea
\lefteqn{\left[|A_\parallel^{\mbox{{\tiny C}}}(t)|^2\right]=
\frac{\gf^2}{4}|V_{ub}V_{cs}|^2
|M_\parallel|^2}\lab{e41}\\
&&\times\left[\left(1+|R_\parallel|^2
\right)\left(e^{-\Gamma_Lt}+e^{-\Gamma_Ht}\right)+2|R_\parallel|
\cos(\rho_\parallel+\gamma)\left(e^{-\Gamma_Lt}-e^{-\Gamma_Ht}\right)
\right]\nonumber
\eea
\bea
\lefteqn{\left[|A_\perp^{\mbox{{\tiny C}}}(t)|^2\right]=
\frac{\gf^2}{4}|V_{ub}V_{cs}|^2|M_\perp|^2}\\
&&\times\left[\left(1+|R_\perp|^2\right)\left(e^{-\Gamma_Lt}+
e^{-\Gamma_Ht}\right)-2|R_\perp|\cos(\rho_\perp+\gamma)
\left(e^{-\Gamma_Lt}-e^{-\Gamma_Ht}\right)\right]\nonumber
\eea
\bea
\lefteqn{\left[A_0^{\mbox{{\tiny C}}\ast}(t)A_\parallel^{\mbox{{\tiny
C}}}(t)\right]=\frac{\gf^2}{4}|V_{ub}V_{cs}|^2|M_0||M_\parallel|
e^{i(\vartheta_\parallel-
\vartheta_0)}\left[\left(1+|R_0||R_\parallel|e^{i(\rho_\parallel-\rho_0)}
\right)\right.}\lab{e43}\\
&&\left.\times\left(e^{-\Gamma_Lt}+e^{-\Gamma_Ht}\right)+\left(
|R_0|e^{-i(\gamma+\rho_0)}+|R_\parallel|e^{i(\gamma+\rho_\parallel)}\right)
\left(e^{-\Gamma_Lt}-e^{-\Gamma_Ht}\right)\right]\nonumber
\eea
\bea
\lefteqn{\left[A_\parallel^{\mbox{{\tiny C}}\ast}(t)
A_\perp^{\mbox{{\tiny C}}}(t)\right]=
\frac{\gf^2}{4}|V_{ub}V_{cs}|^2|M_\parallel||M_\perp|
e^{i(\vartheta_\perp-\vartheta_\parallel)}\left[\left
(1-|R_\parallel||R_\perp|e^{i(\rho_\perp-\rho_\parallel)}
\right)\right.}\\
&&\left.\times\left(e^{-\Gamma_Lt}+e^{-\Gamma_Ht}\right)+\left(
|R_\parallel|e^{-i(\gamma+\rho_\parallel)}-|R_\perp|e^{i(\gamma+
\rho_\perp)}\right)\left(e^{-\Gamma_Lt}-e^{-\Gamma_Ht}\right)\right]\nonumber
\eea
\bea
\lefteqn{\left[A_0^{\mbox{{\tiny C}}\ast}(t)
A_\perp^{\mbox{{\tiny C}}}(t)\right]=
\frac{\gf^2}{4}|V_{ub}V_{cs}|^2|M_0||M_\perp|
e^{i(\vartheta_\perp-\vartheta_0)}\left[\left
(1-|R_0||R_\perp|e^{i(\rho_\perp-\rho_0)}
\right)\right.}\lab{e45}\\
&&\left.\times\left(e^{-\Gamma_Lt}+e^{-\Gamma_Ht}\right)+\left(
|R_0|e^{-i(\gamma+\rho_0)}-|R_\perp|e^{i(\gamma+
\rho_\perp)}\right)\left(e^{-\Gamma_Lt}-
e^{-\Gamma_Ht}\right)\right].\nonumber
\eea
Combining these equations appropriately -- each of them represents a
certain measurement -- a determination of $\gamma$ and of the strong
phase shifts is possible without using any additional input. This can
be seen as follows:

Let us consider the untagged observables corresponding to (\ref{e34}),
(\ref{e35}) and to the real part of (\ref{e37}). From these rates the
{\it ratios} of the coefficients of $e^{-\Gamma_Lt}-e^{-\Gamma_Ht}$ and
of $e^{-\Gamma_Lt}+e^{-\Gamma_Ht}$ can be determined. The overall
normalizations of these rates cancel in the ratios which are given by
\beq\lab{eR1}
u_f\equiv\frac{2|R_f|\cos(\rho_f-\gamma)}{1+|R_f|^2}\qquad\left(
f\in\{\,0\,,\,\parallel\,\}\right)
\eeq
and
\beq\lab{eR2}
u_{0,\parallel}\equiv\frac{|R_0|\cos(\vartheta_\parallel-\vartheta_0
-\rho_0+\gamma)+|R_\parallel|\cos(\vartheta_\parallel-\vartheta_0+
\rho_\parallel-\gamma)}{\cos(\vartheta_\parallel-\vartheta_0)
+|R_0||R_\parallel|\cos(\vartheta_\parallel-\vartheta_0+\rho_\parallel
-\rho_0)},
\eeq
respectively, and depend thus only on $|R_0|$,
$\rho_0$, $|R_\parallel|$, $\rho_\parallel$,
$\vartheta_\parallel-\vartheta_0$ and on the CKM angle $\gamma$.
Using in addition the observables of the untagged $B_s$ decays into
the charge conjugate final states that are related to (\ref{e40}),
(\ref{e41}) and to the real part of (\ref{e43}), we can determine similar
ratios of the coefficients of $e^{-\Gamma_Lt}-e^{-\Gamma_Ht}$ and
$e^{-\Gamma_Lt}+e^{-\Gamma_Ht}$. These charge conjugate ratios, which
are given by
\beq\lab{eR3}
u_f^{\mbox{{\tiny C}}}\equiv\frac{2|R_f|\cos(\rho_f+\gamma)}
{1+|R_f|^2}\qquad\left(f\in\{\,0\,,\,\parallel\,\}\right)
\eeq
and
\beq\lab{eR4}
u_{0,\parallel}^{\mbox{{\tiny C}}}\equiv\frac{|R_0|\cos(\vartheta_\parallel-
\vartheta_0-\rho_0-\gamma)+|R_\parallel|\cos(\vartheta_\parallel-\vartheta_0+
\rho_\parallel+\gamma)}{\cos(\vartheta_\parallel-\vartheta_0)
+|R_0||R_\parallel|\cos(\vartheta_\parallel-\vartheta_0+\rho_\parallel
-\rho_0)},
\eeq
respectively, depend on the same six ``unknowns'' as (\ref{eR1}) and
(\ref{eR2}) determined from (\ref{e34}), (\ref{e35}) and (\ref{e37}).
We have therefore six observables at our disposal to determine
the six ``unknowns'' $|R_0|$, $\rho_0$, $|R_\parallel|$, $\rho_\parallel$,
$\vartheta_\parallel-\vartheta_0$, $\gamma$. In particular
we are in a position to extract the CKM angle $\gamma$. Using furthermore
the observables we have not considered so far, certain discrete ambiguities
are resolved and also $|R_\perp|$, $\rho_\perp$,
$\vartheta_\perp-\vartheta_0$ can be determined. Note that the
overall normalizations of the rates corresponding to (\ref{e34})-(\ref{e45})
inform us about $|V_{ub}V_{cs}|\cdot|M_f|$, where
$f\in\{\,0\,,\,\parallel\,,\,\perp\,\}$.

Obviously the major goal of this approach is the extraction of
the CKM angle $\gamma$. However, also the the quantities $|R_f|$ and the
strong phase shifts $\rho_f$, $\vartheta_f$ are of interest, since they
allow insights into the hadronization dynamics of the corresponding
four-quark operators.

\section{CP violation}\lab{cpviol}

There are many CP-violating observables that can be constructed from tagged
time-dependent measurements. Some of them survive even when only untagged
data samples are used. The most striking {\it untagged} CP-violating
observable is
\bea
\lefteqn{\mbox{Im}\left\{[A_f^\ast(t)\,A_\perp(t)]\right\}+
\mbox{Im}\left\{[A^{\mbox{{\tiny C}}\ast}_f(t)\,A^{\mbox{{\tiny C}}}_\perp(t)]
\right\}=-\frac{\gf^2}{2}|V_{ub}V_{cs}|^2|M_f||M_\perp|}\nonumber\\
&&\times\left\{|R_f|\cos(\rho_f+
\vartheta_f-\vartheta_\perp)+|R_\perp|\cos(\rho_\perp
+\vartheta_\perp-\vartheta_f)\right\}\left(e^{-\Gamma_Lt}-e^{-\Gamma_Ht}
\right)\sin\gamma,\lab{cpvio1}
\eea
where $f\in\{\,0\,,\,\parallel\,\}$. Note that the plus sign on the 
l.h.s.\ of that equation is due to the fact that the parity eigenvalues 
of the final state configurations $f$ and $\perp$ arising in the ``mixed'' 
combinations are different. The CP observable (\ref{cpvio1}) is 
proportional to $\sin\gamma$ and occurs even when all strong phase shifts 
vanish. This CP-violating effect can be potentially very large as can be 
seen by employing the factorization assumption which implies vanishing
strong phase shifts.

In contrast, to observe CP violation in the {\it untagged} interference
term involving final state configurations with equal parity eigenvalues
requires non-vanishing strong phase shifts as can be seen from the
corresponding CP-violating observable
\bea
\lefteqn{\mbox{\,Re\,}\left\{[A_0^\ast(t)\,A_\parallel(t)]\right\}-
\mbox{\,Re\,}\left\{[A^{\mbox{{\tiny C}}\ast}_0(t)\,A^{\mbox{{\tiny
C}}}_\parallel(t)]
\right\}=\frac{\gf^2}{2}|V_{ub}V_{cs}|^2|M_0||M_\parallel|}\nonumber\\
&&\times\left\{|R_0|\sin(\rho_0+\vartheta_0-\vartheta_\parallel)+
|R_\parallel|\sin(\rho_\parallel+\vartheta_\parallel-\vartheta_0)
\right\}\left(e^{-\Gamma_Lt}-e^{-\Gamma_Ht}\right)\sin\gamma.
\eea

The last category of CP-violating effects in {\it untagged} data
samples is related to
\bea
\lefteqn{\left[|A_f(t)|^2\right]-\left[|A^{\mbox{{\tiny
C}}}_f(t)|^2\right]}\nonumber\\
&&=\etap\gf^2|V_{ub}V_{cs}|^2|M_f|^2|R_f|\sin\rho_f
\left(e^{-\Gamma_Lt}-e^{-\Gamma_Ht}\right)\sin\gamma
\eea
with $f\in\{\,0\,,\,\parallel\,,\,\perp\,\}$ and requires also
non-vanishing strong phase shifts. This last category is the
only one that has been considered so far in the literature \cite{dunietz}.

\section{Summary}\lab{sum}

We have calculated the time-dependences of the observables of angular
distributions for $B_s$ decays caused by $\bar b\to\bar cu\bar s$
quark-level transitions both in a general notation and in terms of
linear polarization states. Examples for exclusive modes belonging
to this decay category are the color-allowed and color-suppressed 
channels $\bstack\to D_s^{\ast\pm}\, K^{\ast\mp}$, $D_{s1}(2536)^\pm 
K^{\ast\mp}$, $D_s^{\ast\ast\pm}K^{\ast\mp}$ and $\bstack\to\dstackoast\phi$, 
$\dstackres\phi$, $\dstackresres\phi$, respectively. Since charged particles 
are easier to detect for generic detectors than the photon(s) in the 
strong or electromagnetic decays of $D_s^\ast$ and $D^{\ast0}$, we have 
also listed higher resonances exhibiting significant all-charged final 
states.  The information that is provided by the corresponding angular 
correlations allows -- without any theoretical input --  the extraction 
both of the notoriously difficult to measure CKM angle $\gamma$ and of 
the whole hadronization dynamics of these decays thereby allowing e.g.\ 
tests of the factorization hypothesis.

If the lifetime difference between the $B_s$ mass eigenstates $B_s^L$
and $B_s^H$ is sizable, as is indicated by certain present theoretical
analyses, even {\it untagged} $B_s$ data samples suffice to accomplish
this ambitious task. Interestingly, some of the many CP-violating
observables that can be constructed from tagged measurements survive also
in that {\it untagged} case and are potentially very large. One class of
these untagged CP-violating observables is proportional to $\sin\gamma$
and arises even when all strong phase shifts vanish.

{}From an experimental point of view, untagged analyses of $B_s$-meson
decays are obviously much more efficient than tagged studies. The
feasibility of our {\it untagged} strategies for extracting $\gamma$ in
a clean way depends, however, crucially on a sizable lifetime difference
of the $B_s$ system. Even if this lifetime splitting should turn out
to be too small for untagged analyses, once a non-vanishing
lifetime difference has been established experimentally, the formalism
presented in our paper must be used in the case of tagged measurements
in order to extract $\gamma$ correctly. Clearly time will tell and an
exciting future may lie ahead of us.

\section*{Acknowledgments}

We are very grateful to Helen Quinn for a critical reading of the
manuscript. R.F.\ would like to thank Helen Quinn for interesting
discussions and the Theoretical Physics Groups of Fermilab
and SLAC for the warm hospitality during parts of these investigations.
This work has been supported in part by the Department of Energy,
Contract No.\ DE-AC02-76CH03000.

\end{document}